\documentclass[%
 aip,
 amsmath,amssymb,
 reprint,%
]{revtex4-1}

\usepackage{graphicx}% Include figure files
\usepackage{dcolumn}% Align table columns on decimal point
\usepackage{bm}% bold math
\usepackage[utf8]{inputenc}
\usepackage[T1]{fontenc}
\usepackage{mathptmx}
\usepackage{etoolbox}

\makeatletter
\def\@email#1#2{%
 \endgroup
 \patchcmd{\titleblock@produce}
  {\frontmatter@RRAPformat}
  {\frontmatter@RRAPformat{\produce@RRAP{*#1\href{mailto:#2}{#2}}}\frontmatter@RRAPformat}
  {}{}
}%
\makeatother
\begin{document}

\preprint{AIP/123-QED}

\title[Near Kelvin Temperature Phonon Antibunching using Carbon Nanotubes ]{Near Kelvin Temperature Phonon Antibunching using Carbon Nanotubes}
% Force line breaks with \\
\author{Sai Subramanian B}
 \affiliation{Department of Physics, Indian Institute of Madras, Chennai 600036, India}%Lines break automatically or can be forced with \\

\author{Prabhu Rajagopal}
\email{prajagopal@iitm.ac.in}
\affiliation{%
Centre for Nondestructive Evaluation and Department of Mechanical Engineering, Indian Institute of Technology Madras, Chennai, 600036, India.%\\This line break forced% with \\
}%

\date{\today}% It is always \today, today,

\begin{abstract}{\centering {ABSTRACT}\\}
This paper discusses the development of a novel setup to achieve Phonon Antibunching at near Kelvin temperatures using carbon nanotubes. A previously experimentally studied electro-mechanical single carbon nanotube setup is proposed to be modified by the addition of another coupled nanotube. This modified set-up is then described analytically, and low temperature approximations to the mechanical energy are used to obtain the quantum Hamiltonian. The stationary Liouville-Von Neumann master equation is then used to demonstrate antibunching at near Kelvin temperatures using the proposed design. The achievement of phonon antibunching at such large temperatures suggests a deeper study into the use of carbon nanostructures in this field. It also brings the possibility of using phonon antibunching for quantum computing and sensors a step closer to reality.
\end{abstract}

\maketitle

\section{\label{sec:level1}Introduction\protect  }

Phonons quantify lattice vibrations\cite{intro,intro2}, and phononics is a rapidly emerging field leveraging recent advances in elastic/mechanical wave phenomena including diverse branches such as acoustics, ultrasonics and thermal transport. The study of mechanical waves at low Temperatures up to a few Kelvin where quantization becomes important, informs quantum phononics, with applications in computing, sensing and diagnostics\cite{app1}.

Today photonics is typically the field of choice for physical realizations of quantum information theory. The reasons for this include great speed, environmental stability, and good integrability. Quantum phononics on the other hand has many other benefits including longer stability to damping, and the possibility of interaction with a wide range of other quantum systems including electric, magnetic, and optical systems. Quantum phononics can be conducted in a larger range of materials than its photonic counterparts. These different materials can also be cast in the form of different waveguides each with their own unique set of properties. Therefore Quantum phononics is also an excellent candidate for quantum information carriers\cite{Qc,Qc1,Qc2}. With a large amount of research interest, many workers are also exploring phononic quantum computing as a vital alternative in the 'quantum computing basket'.

Quantum phononics generally works in relatively large time scales and it also has a very good depth of penetration in various media. Moreover, since it is based on elastic waves, phononics is typically non-irradiative unlike its electromagnetic wave based counterparts (such as photonics or electronics). This make quantum phononics an excellent option for sensing applications\cite{Sn,Sn1}. Another advantage of using quantum phononics is the exciting possibility of recycling of waste heat and energy to create on-demand or ‘point-of-care’ solutions. This could revolutionize the fields of sensing and quantum computing. To make these technologies a reality and to try and rival the success of quantum photonics with quantum phononics, we first need to achieve phonon antibunching. Being Bosonic in nature, special mechanisms are needed to prevent the co-existence of multiple phonon energy states in the quantum harmonic oscillator understanding of wave mechanics. Several analytical and simulation studies have proposed approaches to achieve phonon antibunching \cite{An1,AnC1,AnC2,AnC3,Guan,Pr21,AnU1,AnU2}.

 Most of these studies have been focused on replicating ideas from photon antibunching in the phononic context. One key idea is to create a very large non-linearity in the system, thereby causing the second and higher excited states of the quantum harmonic oscillator to become very unstable. This is called the 'conventional' phonon/photon/Coulomb blockade depending on if the system is phononic/photonic or electronic \cite{AnC1,AnC2,AnC3}. However most systems do not support such high non-linearity. A second idea was thus developed which made use of coupling along with a weak non-linearity. This is called the unconventional photon/phonon blockade \cite{Guan,Pr21,AnU1,AnU2}. The problem with the unconventional phonon blockade is that unlike in photonics, typically frequencies are relatively smaller, and thus we have to contend with thermal fluctuations at near-Kelvin temperatures. We therefore have to work with the system at millikelvin temperatures to achieve phonon antibunching, imposing severe demands on cooling needed. 

Carbon nanostructures have an extremely diverse range of waveguides each with their own unique properties \cite{cn,cn2}. They have helped create advances in several fields including electronics, optics, nanotechnology and material science \cite{cna}. Here, we propose their use for several reasons including availability of high frequency waveguides, possibility of relatively high non-linearity among other things\cite{Laird39,cne2,cn,cne3}. They have also been extensively studied experimentally and can easily be fabricated for commercial use. Theoretical studies so far seem to be predominantly classical and a complete quantum analysis in the required temperature regime for many of the unique waveguides is hard to find.

An experimental study of a nano-electromechanical  carbon nanotube based system performed near the quantum ground state demonstrated high frequencies up to 37 GHz \cite{Laird39}. Here we propose to modify this setup to include coupling and develop anti-bunching. This system has been completely analysed classically \cite{CNTD,CNTD2}. We start with the classical analysis and then make the required low temperature and quantum approximations to reach the quantum Hamiltonian. We then simulate the Hamiltonian to find the solution states and checked for antibunching at up to near kelvin temperatures. We explain our results including finding antibunching at up to near-Kelvin temperatures including possible reasons for the marked improvement in temperature performance. We conclude that this improvement warrants more research into the use of carbon nanostructures and their unique waveguides for realizing quantum phononics.

\section{Model}
Fig 1 depicts the model studied here. Two carbon nanotubes with identical geometrical and mechanical properties (radius r, length L and elastic modulus E) are suspended, one just above the other. They are clamped on both ends to metal pads. Without loss of generality, we can take the voltages to be V,0 on the left and right side respectively. They are suspended over a gate with Voltage $V_G$. The height of the pads from the gate is R. Let the capacitive coupling between the gate and the nanotubes be $C(z)$.There will also be a capacitive coupling with the metal pads given by $C_L(z)$ and $C_R(z)$ respectively on the left and right. The axes directions are also defined in Fig 1.

Our model was partly inspired by the experimental achievement of high frequency modes up to 37 GHz with a similar model in (Ref.~\onlinecite{Laird39}). However (Ref.~\onlinecite{Laird39}) had studied only a single carbon nanotube. Our modification to add a second nanotube seemed very natural given most approaches to achieve anti-bunching so far have had a large focus on creating coupling between two systems. 

\begin{figure}
\includegraphics[width = 1.2\linewidth]{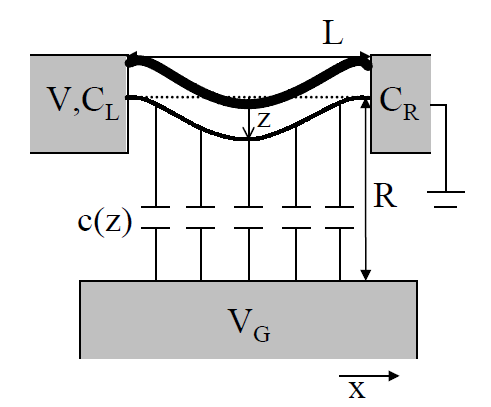}
\caption{\label{fig:setup}Schematic Diagram of our model, modified from the setup analyzed in (Ref.~\onlinecite{CNTD}) and experimentally studied in (Ref.~\onlinecite{Laird39})}
\end{figure}

\subsection{Theoretical Analysis}

We begin with classical calculations done for a single carbon nanotube in this setup taken from (Ref.~\onlinecite{CNTD}). The total energy in the system is the sum of the elastic and electrostatic energy given in Equations (1) and (2) .

    Elastic Energy 
        \begin{equation}
            W_m(z(x)) = \int_{0}^{L}[\frac{EI}{2}z"^2 + T z'^2] dx
        \end{equation}
        where $ S = \pi r^2 $ , $ I = \frac{\pi r^4}{4} $ and $ T = \frac{ES}{8L}\int_{0}^{L}z'^2 dx$ and ' is derivative with respect to x.
 
    Electrostatic Energy
  \begin{equation}
            W_e(z(x)) = -K_0\int_{0}^{L}z(x) dx
       \end{equation}
       where $ K_0 = \frac{(ne)^2}{L^2R} $ and ne is the total charge.
    
       We get the solution z(x) by minimizing the energy. The solution is given in Equation (3).
       
     \begin{equation}
            z(x) = \frac{K_0L}{2T\epsilon}[\frac{sinh \epsilon L}{cosh \epsilon x -1}(cosh \epsilon x -1) - sinh \epsilon x + \epsilon x - \frac{\epsilon x^2}{L} ]
       \end{equation}
       where $ \epsilon = \sqrt{\frac{T}{EI}}$ and $ T = \frac{K_0 L^6 S}{60480 EI^2}$   
       
        We are working under the low temperature limit and therefore we can assume $\epsilon x << 1 $, we use that to simplify the expression for z(x).
    
     \begin{equation}
            z(x) = \frac{K_0x^2}{T} ( \frac{1}{2} + \frac{\epsilon L}{4}\frac{sinh \epsilon L}{cosh \epsilon L -1})
       \end{equation}
      The approximation seems to hold good and match what we would physically expect for $x<L/2$.

       The value of z' and z" becomes,
       \begin{equation}
       z' =  \frac{2z}{x}
       \end{equation}
       \begin{equation}
            z" = k
       \end{equation}
       where k is a constant.
      
    We now apply these approximations to the energy expressions (Equations (1) and (2)). We are looking to study small oscillations about the steady state solution and therefore need to also include a Kinetic Energy term $W_k$  given by Equation (9).
    
    \begin{equation}
      W_m(z(x,t)) =  T \int_{0}^{L}(\frac{2z(x)}{x})^2 dx
    \end{equation}
     \begin{equation}
         W_e(z(x,t)) = K_0 \int_{0}^{L}z(x) dx
     \end{equation}
    \begin{equation}
        W_k(z(x,t)) = \frac{m}{2L} \int_{0}^{L}(\frac{dz(x)}{dt})^2 dx
    \end{equation}
   
  Our goal is to be able to write the energy as the Quantum Hamiltonian, to use the Ladder Operator Formalism, we make one last approximation. We approximate the integrals by their value at $L/2$. This approximation is justified in the Temperature scales we are working in.
   
   Making this approximation, we can rewrite the energy expressions (Equations (7) to (9)) as Equations (10) to (12),
   
   \begin{equation}
      W_m =  (\frac{4Tz(L/2)}{L})^2
    \end{equation}
     \begin{equation}
         W_e = K_0 z(L/2)
     \end{equation}
    \begin{equation}
        W_k = \frac{m}{2L} (\frac{dz(L/2)}{dt})^2
    \end{equation}
    
    We are now ready to define our ladder operators for the variable $z(L/2)$ (Equations (13) and (14)). From now on we just refer to it as z.
    
    \begin{equation}
    b_+ = \frac{1}{\sqrt{(2 \hbar m \omega)} }(-im \frac{dz}{dt} + m\omega z)
\end{equation}
\begin{equation}
    b_- = \frac{1}{\sqrt{(2 \hbar m \omega)} }(+im \frac{dz}{dt} + m\omega z)
\end{equation}

where $\omega^2 = \frac{32T}{mL}$

We can now write the energies (Equations (10) to (12)) as terms in the Quantum Hamiltonian given by Equations (15) and (16),

\begin{equation}
    W_m + W_k = \hbar \omega (b_+ b_- + \frac{1}{2})
\end{equation}
\begin{equation}
    W_e = F (b_+ + b_-)
\end{equation}
where F is defined as the forcing constant.

So far we have treated T as a constant but actually it does vary with z'. We will now accommodate that by adding an extra term,

\begin{equation}
    T = \frac{ES}{8L}\int_{0}^{L}z'^2 dx
\end{equation}

The Elastic Potential energy is proportional to $Tz^2$, so the term we introduce should be proportional to $z^4$ given by Equation (18).

\begin{equation}
    U(b_+ + b_-)^4
\end{equation}

We make the rotating wave approximation to get Equation (19),

\begin{equation}
    U(b_+^2 b_-^2)
\end{equation}

where U is the non-linear constant

Each carbon nanotube in our system will have the above terms. Along with that, there will be a coupling energy between the two carbon nanotubes. The coupling energy will be proportional to the displacement of both carbon nanotubes and is given by Equation (20).

\begin{equation}
    J(b_+ + b_-)(d_+ + d_-)
\end{equation}

where J is the coupling constant. We once again make the rotating wave approximation to get Equation (21).

\begin{equation}
  J(b_+ d_- + b_- d_+)  
\end{equation}

The value of J can be calculated as a function of distance between the carbon nanotubes \cite{Pr21}. We vary the distance in the simulations and take the best possible value of J.
\\

We now put together all the terms to write down our complete Hamiltonian,ignoring constants.

\begin{widetext}
\begin{equation}
 H = \hbar \omega_1 (b_+ b_-) +\hbar \omega_2 (d_+ d_-) + F_1(b_+ + b_-) + F_2(d_+ + d_-) \nonumber\\ + U_1(b_+^2 b_-^2) + U_2(d_+^2 d_-^2) + J(b_+ d_- + b_- d_+) \;. \label{eq:wideeq}
 \end{equation}
\end{widetext}

The Hamiltonian has the standard quantum harmonic oscillator energy terms along with forcing terms, non-linear terms and a coupling term.

\subsection{Quantum Simulation}

Having obtained our Hamiltonian, we now need to need to find the solution state vector and check for anti-bunching. The solution we require can be obtained at finite Temperature as a density matrix from solving the Liouville-Von Neumann Master Equation (Equation 22).

\begin{equation}
    \frac{d\rho}{dt} = L\rho = -i[H,\rho] + \sum_{n=1,2}\gamma/2[(nth+1)D[b_-]\rho +(nth) D[b_+]\rho]
\end{equation}

where $D[A]\rho = 2A\rho A^\dag - A^\dag A \rho - \rho A^\dag A$ and $nth = 1/(e^{T/T_0}-1)$ and $T_0 = \hbar \omega/K_b$
\\

 The magnitude of antibunching is measured by the second-order correlation function given by Equation (23). Values of $g_2(0)$ close to zero indicate anti-bunching.
 
    \begin{equation}
        g_2(0) = \frac{Tr(b_+b_+b_-b_- \rho)}{Tr(b_+b_- \rho)^2}
    \end{equation}
    
We will be working with temperatures not more than one or two Kelvin. The number of phonons can therefore be limited to single digits. The differential Equation (22) to find the state vector can be written as a matrix eigenvalue equation. This can be solved on a computer using numerical methods in commercial codes \cite{num1,Pr21}.
\section{Results}

We have run simulations on various possible values of the variable parameters. Experimental evidence from (Ref.~\onlinecite{Laird39}) tells us that this kind of system can support higher frequencies in the range of tens of GHz. Higher the frequency, larger the temperature at which we can achieve anti-bunching. Therefore we have run our simulations with 37 GHz as the frequency of both resonators. We varied the other parameters throughout their possible physical ranges. In figs 2 and 3 we can see the graphs showing the variation of $Log (g_2)$ with respect to the some of the important parameters suitably scaled to study variation.
\\

Fig 2 shows the variation of $Log (g_2(0))$ with J for various $F_1$ values. This was the parameter we had maximum freedom to vary. Our results are in line with those achieved using unconventional phonon blockade and we can see clearly that there is no antibunching without coupling. We can also see that while the forcing term is necessary, increasing it too much is detrimental to antibunching. 
\\

Fig 3 shows the variation of $Log (g_2(0))$ with $U_2$ for various $U_1$ values.A linear system is expected to have no anti-bunching. This was also verified during the simulations. The other positive feature of our proposed system is that only weak non-linearity which is more physically possible is needed.
\\

The results in Figs 2 and 3 tell us how well carbon nanostructures and their unique waveguides can fit into the current method of study of phonon antibunching with some proper quantum mechanical analysis. Fig 4 tells us why we should include them in this field. At very low temperatures we were able to achieve a $g_2(0)$ value near $10^{-11}$. This is orders of magnitude better than those shown in earlier studies. We are able to maintain anti-bunching up to Temperatures near 0.1 K. This is again an order of magnitude improvement on the millikelvin temperature anti-bunching achieved earlier.

\begin{figure}
\includegraphics[width = 1.1\linewidth]{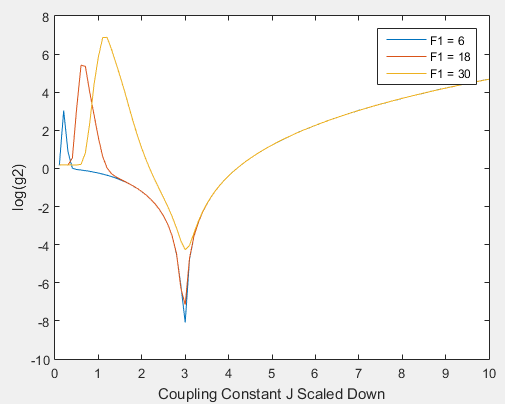}
\caption{\label{fig:JF}Plot of $Log (g_2(0))$ vs J/20}
\end{figure}

\begin{figure}
\includegraphics[width = 1.1\linewidth]{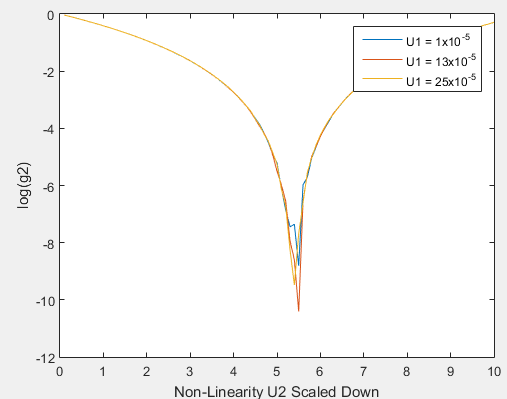}
\caption{\label{fig:UU}Plot of $Log (g_2(0))$ vs $U_2$/$10^{-5}$ }
\end{figure}

\begin{figure}
\includegraphics[width = 1.0\linewidth]{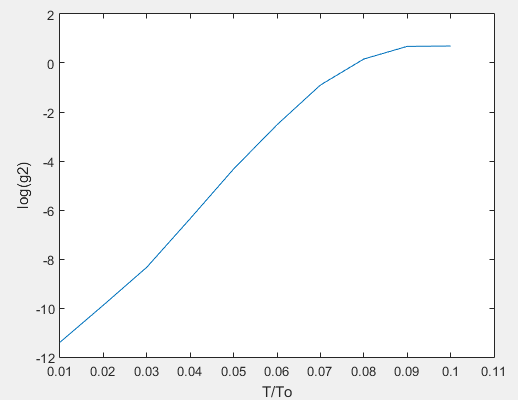}
\caption{\label{fig:T}Plot of $Log (g_2(0))$ vs $T/T_0$ }
\end{figure}

\section{Discussion}

There are several reasons as to how we could achieve a raise in the temperature of the system up to 0.1K while still maintaining antibunching, considering the fact that most of the work in this field has been in the milliKevin temperature zone. The first is the fact that our study considered very high frequencies. The influence of thermal phonons at a given temperature depend on their energy which in turn depend on their frequency. Larger frequencies imply larger energies which imply higher temperatures for thermal phonons to become important. Another major advantage of using this system was in the number of parameters available to vary and the greater freedom available in varying them. 
\\

Through the variation of all the parameters we were able to continually see results we would expect physically. This is particularly true at the extreme values of the parameters. This was quite important to us as this is the first time waveguides like those of the carbon nanotubes were used in a phonon antibunching study. It gives us more confidence in the validity of our approach in the quantum mechanical analysis as well as in the approximations we chose to make.
\section{Conclusions and Future Work}

In this paper, we introduced a new setup using carbon nanotubes to develop phonon antibunching. We began with the classical energy equations and made low-temperature simplifications. We were then able to write them in the form of a proper quantum mechanical Hamiltonian in the language of the ladder operators. We simulated the the Hamiltonian to find the solution state and check for anti-bunching. We were able to improve upon the anti-bunching results by order of magnitude achieving antibunching at temperatures near 0.1K. We also saw that the results for different values of the parameters were matching physical expectations hinting that the approach and approximations are justified. We finally looked at possible reasons why our carbon nanotube setup were able to achieve these results.
\\

We can conclude that the method used for quantum mechanical analysis was successful. There are many more waveguide embodinments possible using carbon nanostructures. Each of them have their own set of unique properties. One could definitely apply the method used in this study to develop new setups and analyze them to search for antibunching. Studying the ranges of the parameters where there is antibunching could also help us better understand this phenomenon. We also hope to see this setup tested experimentally.
\\

Our team is working on developing an experimental setup to test the finding of this paper. We are also working on developing other setups including possibly coupling of carbon nanostructure resonators with optical or other modes.We believe that these results contribute to bringing the field of quantum phononics closer to real and commercial applications.

\section*{Data Availability Statement}

The data that support the findings of this study are available from the corresponding author upon reasonable request.

\section*{References}

\nocite{*}
\bibliography{aipsamp}% Produces the bibliography via BibTeX.

\end{document}